# Report on article The Travelling Salesman Problem: A Linear Programming Formulation

Radosław Hofman, *Poznań 2008*

*Abstract*—This article describes counter example prepared in order to prove that linear formulation of TSP problem proposed in [7] is incorrect (it applies also to QAP problem formulation in [8]). Article refers not only to model itself, but also to ability of extension of proposed model to be correct.

## I. INTRODUCTION

Unknown relation between P and NP [3] complexity classes remains to be one of significant non solved problems in complexity theory. P complexity class consists of problems solvable by Deterministic Turing Machine (DTM) in polynomially bounded time, while NP complexity class consists of problem solvable by Non Deterministic Turing Machine (NDTM) in polynomially bounded time. This means that DTM can verify solution of every NP problem in polynomially bounded time even if polynomial algorithm for finding this solution is unknown [16].

Significant subclass of NP problems is known as NP-complete class. Problems from this class have ability to represent any other problems from whole NP complexity class. In 1970 S. Cook presented in [2] first reduction from any NP problem to Boolean Satisfiability Problem, and two years after R. Karp proved that 21 other problems are in NP-complete class showing many-one polynomial time reductions to these problems [14]. If then anyone shows algorithm solving any NP-complete problem in polynomially bounded time then any of NP problems may be solved in no more then $O(n^c)$ steps, where $n$ stands for instance size and $c$ is some constant value [16].

In 2006 there appeared articles claiming to have proven that P=NP formulating TSP and QAP problem in terms of linear programming ([4], [10], [5] and other). Author of this article have prepared counter example [11] for one of these article discussing inability for LP approach to solve large instances of NP-complete problems [12].

Some suggestions from [11] were taken into proposed model and counter example is not valid for new version of these models. In this article we present extended version proving flaws in extended model.

## II. MODEL LIMITATIONS

### A. 3D space

Model consists of variables representing usage of flow between nodes $i$ and $j$ at some stage $r$. Example is presented below:

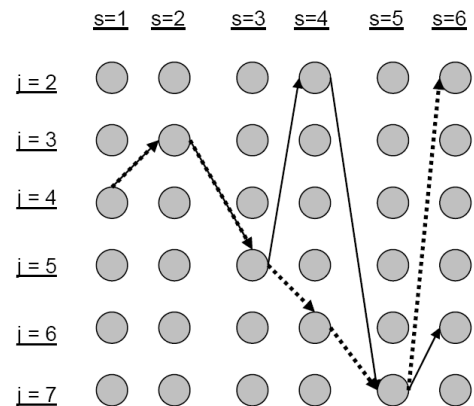

Figure 1 Example flow

Using these 3D variables (there are $O(n^3)$ variables where $n$ is number of nodes) for linear formulation could not prevent of solutions where:
- flow splits to clusters
- within clusters flow reaches same nodes several times
- summary flow at each node is preserved

Example of flow beyond restrictions for 3D variables is presented at fig. 2.

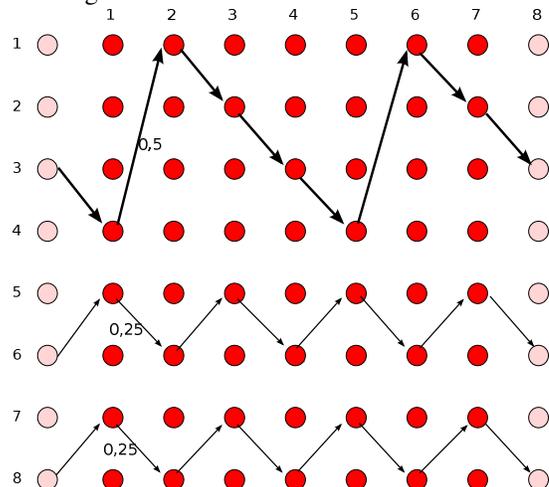

Figure 2 Flaws in LP model based on 3D variables

Of course these 3D variables would be sufficient for Integer Linear programming, but IP is known to be NP-complete [14].

*B. 6D space*

Model may be extended to 6D variables. In fact one might say that it is 3D x 3D – for every 3D variable we build whole solution.

It may seem as easy task to find appropriate graph but below consideration is a result of months of experiments.

First thing, as observed in Usenet by David Moews, and in [11] probable counter example for whole model requires more then 50 cities. Verification of optimal solution or generation of sole variables of discussed model is out of reach for standard computers in rational time and space. Building counter example was then based on instances for HCP assuming that we will assign small cost for each arc in HCP instance and large cost otherwise. Proposing solution we had also to invent way to enlarge graph without any change to optimal solution. We had used 2 possible enlargements: if node is coincident with 2 arcs then it may be replaced with 2 nodes, and if is coincident with 3 arcs then may be replaced with 3 nodes as presented on fig 3.

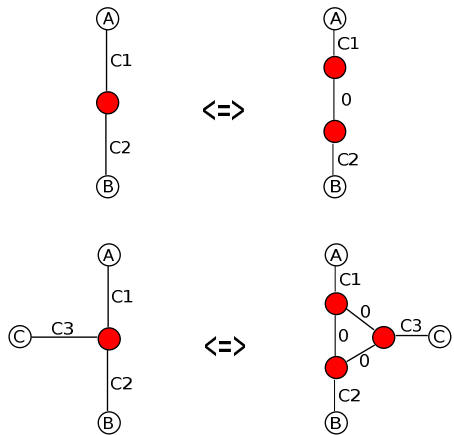

Figure 3 Enlargements not changing optimal TSP tour

It is obvious that such replacements does not change TSP tour and optimal solution value (in first route from A⇔B can be used only once and cost is C1+C2, in second route from A⇔B can be used once and it prevents usage of routes A⇔C and B⇔C with same cost as in original solution, analogically when A⇔C is chosen then A⇔B nor C⇔B may be used and so on). Of course such enlargement cannot be applied for node coincident with more then 3 arcs, because it changes optimal solution (in new graph selection of one route does not prevent of using another one).

Further on we will use name "Replacement nodes" to point out that we give information about new pair or triple of nodes, and "external flow" to address arcs coincident with "replacement nodes" but not arcs between them.

Then we have constructed HCP instance where each node has at most 3 arcs. Our instance containing 23 nodes is presented on figure 4. This instance answer is "NO" – there is no Hamiltonian Cycle in graph. After transformation to TSP instance we obtain optimal solution 19*[small cost]+3*[large cost] (there are more then 2.000.000 such solutions).

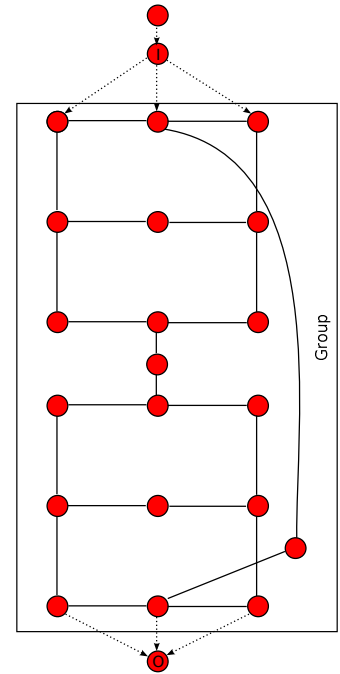

Figure 4 HCP instance, there is one direction link between node "I" and each node in "Group", and from each node in "Group" to node "O"

Next step is to replace each node in "Group" using appropriate pattern presented on fig 3 (we take into account only arcs coincident to 2 nodes in "Group", so arcs form "I" and to "O" are not considered here). As an result we do get instance containing 51 nodes (48 in "Group") with optimal solution containing 3*[large cost].

Our solution (assignment values to variables) is constructed using below rules:
- flow from "I" is splited to 48 parts, each to different node in "Group"
- at each following step flow from each node is splited
    o 0,5 of 1/48 to "external flow" (original arc)
    o 0,25 of 1/48 from new node to to another "replacement node"
- after 47 such steps there is flow from each node to "O"

Our solution does not include [large cost] so is smaller then optimal TSP tour. In section III.B we will consider restrictions made to original model showing that it is correct solution for discussed linear formulation while incorrect answer to TSP question. We add here only information that for every node in "Group":
- 1/48 of flow is entering each node 48 times
- 1/48 of flow is leaving each node 48 times
- whole flow at node leaves node in each stage

*C. 9D space*

9D space is nothing more then 3D x 6D space – we have to build graph of flows as in this counter example for each 3D variable. In other words for every pair of arcs we have to build complete solution, but again this solution (for chosen pair of arcs) is prone to loops presented on fig. 2. Rules of construction are similar, but we do not consider it here, as what will be shown below model is in fact 6D model.

It is important for reader of this article to understand why 3D model is not sufficient for large instance, and then why 6D model is still insufficient. We may consider 9D, 12D etc models, but adding more dimensions complicates model rising its ability to give correct solutions while it is still not correct for infinitively large instance.

### III. COUNTER EXAMPLE

#### A. Discussed model implementation limitations

In [7] author builds 9D model for TSP problem. In fact he uses only $z_{*,1,*<6D>}$ and $y_{<6D>}$ variables, what makes this model $O(n^6)$ (we assume that flow at stage 1 is known, and in this case $z_{1,1,2,i,s,j,k,r,t}=y_{i,s,j,k,r,t}$). In other words, if one adds one node then model addresses only $(n+1)^6$ variables.

#### B. Model restrictions

Now we will briefly explain why above example cover every equation presented in [7].

*1) Equation 6*

This equation checks if flow for 3 levels is preserved. It is equivalent to pair of equations:
- flow for pairs at stages 1 and 2 is equal to 1
- flow for pairs at stages 2 and 3 is preserved

Obviously preserved.

*2) Equations 7 and 8*

This equation checks for each flow if in sub-graph there is flow conservation (incoming flow is equal to outgoing flow at every node). Equation 7 checks it for stages following selected arc and 8 checks it for stages preceding selected arc.

This restriction is also preserved. For every arc there can be built whole graph of flow.

*3) Equation 9 and 10*

Equations 9 and 10 checks if for selected flow it is equal for each other stages (equation 9 for following stages and equation 10 for preceding stages).

Obviously these restrictions are met basing on the same reasons as above – if complete and consistent graph can be built for each node then it contains same value of flow at each stage.

*4) Equation 11*

This equation checks if for chosen arc flow reaches every node with same flow value. This equation was suggested to author of model in [11], now its added but as stated in that report it does not change ability of model to solve NP-complete problem. Addition of this equation has brought most complications to construction of counter example. We may express it getting more into details using analysis of any possible subgraph containing considered flow:

- it means that for each flow in this subgraph there has to be possibility to "leave" subgraph before flow will enter the same node more then once
  this one is met – for any chosen subgraph "outgoing" flow is greater then $2*0,5*1/48=1/48$; this means that if considered subgraph has $p$ nodes, then at every stage it contains $p*1/48$ of flow, and in $p$ steps at least $p*1/48$ of flow may "leave" this subgraph, so there exist assignment to variables such that it will be fulfilled

- there has to exist path for each arc to every other node independent for nodes used in this arc
  in this graph there are at least two independent paths from each node to every other node in "Group" it means that if we consider selection of an arc, there is still at least one path left to reach every other node and whole flow may be constructed

*5) Equation 12*

Equation 12 is consequence of introducing $z$ variables. For counter example $z_{1,1,2<vector>}=y_{<vector>}$, so obviously restriction is fulfilled.

*6) Equation 13 and 14*

These equations restrict occurrence of invalid variables.

### IV. SUMMARY

In summary we have to stress out that these article presents counter example for method presented in [7] and [8]. Counter example is impossible to be directly calculated, but careful analytic consideration proves its correctness.

Why LP method fails for large instance? One may think about considered polytope as about set of $O(n!)$ vertexes (see fig. 5). If someone tries to express boundaries using less then $O(n!)$ restrictions for those vertexes then first of all, he has to prove that it is possible (that vertexes are organized in $O(n^c)$ facets).

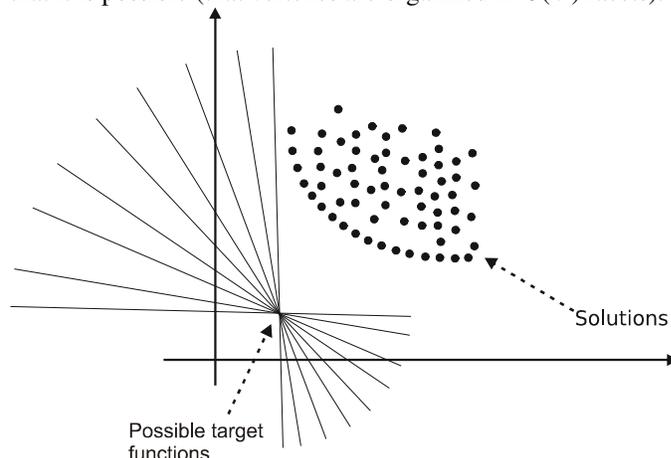

Figure 5 Solutions and possible target functions

Unless such proof is presented then one could not expect that solution found using boundaries has different target value than each of correct solutions and thus cannot be expressed as linear combination of original vertexes (see fig. 6).

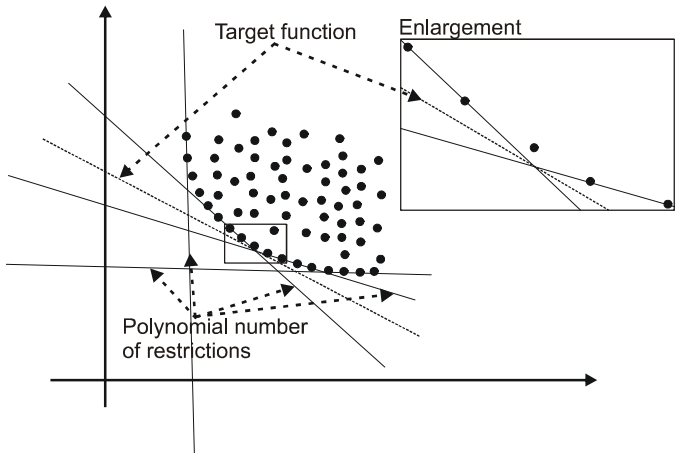

Figure 6 Limited numbers of line restrictions and target function

In summary we also add that discussed model is symmetric despite of authors claims, and because of arguments presented in [17] is theoretically incorrect. It is obvious that $x_{u,p,v}$ variables are building symmetric space. For $y_{u,p,v,k,s,t}$ author adds restriction that $p<s$, but in equations 7-11 he treats them as they were symmetric. Especially in 11: it is obvious that for selected $<u,p,v>$ arc $<k,s,t>$ flows are checked for $s<p$ and $s>p$. If then whole model was presented without restrictions that in $y_{u,p,v,k,s,t}$ $p<s$ then addition of restrictions that $y_{u,p,v,k,s,t} = y_{k,s,t,u,p,v}$ would give exactly the same model. The same consideration applies to $z$ variables. If then one removes half of variables but still uses them only with different notation then it does not change the fact that the model is symmetric.

## REFERENCES

[1] Bazaraa, M.S., Jarvis J.J., Sherali H.D., "Linear Programming and Network Flows", Wiley, New York, 1990
[2] Cook S.A., "The complexity of theorem-proving procedures", Proceedings of the third annual ACM symposium on Theory of computing, 1971, pp. 151-158
[3] Cook S.A., "P versus NP problem", unpublished, Available: http://www.claymath.org/millennium/P_vs_NP/Official_Problem_Description.pdf
[4] Diaby M., "P = NP: Linear programming formulation of the traveling salesman problem.", 2006, unpublished, Available: http://arxiv.org/abs/cs.CC/0609005
[5] Diaby M., "On the Equality of Complexity Classes P and NP: Linear Programming Formulation of the Quadratic Assignment Problem.", Proceedings of the International Multi Conference of Engineers and Computer Scientists 2006, IMECS '06, June 20-22, 2006, Hong Kong, China, ISBN 988-98671-3-3
[6] Diaby M., The Travelling Salesman Problem: A Linear Programming Formulation, WSEAS Transactions on Mathematics 6:6, 2007
[7] Diaby M., A $O(n^8) \times O(n^7)$ Linear Programming Model of Travelling Salesman Problem, unpublished, Available: http://arxiv.org/abs/0803.4354
[8] Diaby M., A $O(n^8) \times O(n^7)$ Linear Programming Model of Quadratic Assignment Problem, unpublished, Available: http://arxiv.org/abs/0802.4307
[9] Evans J.R., Minieka E., "Optimization Algorithms for Networks and Graphs", second ed., Dekker, New York, 1992, pp. 250-267
[10] Gubin S., "A Polynomial Time Algorithm for The Traveling Salesman Problem", 2006, unpublished, Available: http://arxiv.org/abs/cs.DM/0610042
[11] Hofman R., "Report on article: P=NP Linear programming formulation of the Traveling Salesman Problem", 2006, unpublished, Available: http://arxiv.org/abs/cs.CC/0610125
[12] Hofman R., "Why LP cannot solve large instances of NP-complete problems", Proceedings of the International Multi Conference of Engineers and Computer Scientists 2007, IMECS '07, March 2007, Hong Kong, China, ISBN 978-988-98671-4-0,
[13] Karmarkar, N., "A new polynomial-time algorithm for linear programming," Combinatorica 4, (1984) pp. 373-395
[14] Karp R. M., "Reducibility Among Combinatorial Problems.", In Complexity of Computer Computations, Proc. Sympos. IBM Thomas J. Watson Res. Center, Yorktown Heights, N.Y. New York: Plenum, p.85-103. 1972.
[15] Khachiyan, L.G., "Polynomial algorithm in linear programming,", Soviet Mathematics Doklady 20, (1979) pp. 191-194.
[16] Papadimitriou, C.H., Steiglitz K., "Combinatorial Optimization: Algorithms and Complexity", Prentice-Hall, Englewood Cliffs, 1982
[17] Yannakakis, M., "Expressing Combinatorial Optimization Problems by Linear Programs", Journal of Computer and System Sciences 43 (1991) pp. 441-466

## V. ANNEX

### A. Cost table for counter example

Flow from 1:
$cost{1}{2}=1;

Flow from "I":
$cost{2}{3}=1; $cost{2}{4}=1; $cost{2}{5}=1; $cost{2}{6}=1; $cost{2}{7}=1; $cost{2}{8}=1; $cost{2}{9}=1; $cost{2}{10}=1; $cost{2}{11}=1; $cost{2}{12}=1; $cost{2}{13}=1; $cost{2}{14}=1; $cost{2}{15}=1; $cost{2}{16}=1; $cost{2}{17}=1; $cost{2}{18}=1; $cost{2}{19}=1; $cost{2}{20}=1; $cost{2}{21}=1; $cost{2}{22}=1; $cost{2}{23}=1; $cost{2}{24}=1; $cost{2}{25}=1; $cost{2}{26}=1; $cost{2}{27}=1; $cost{2}{28}=1; $cost{2}{29}=1; $cost{2}{30}=1; $cost{2}{31}=1; $cost{2}{32}=1; $cost{2}{33}=1; $cost{2}{34}=1; $cost{2}{35}=1; $cost{2}{36}=1; $cost{2}{37}=1; $cost{2}{38}=1; $cost{2}{39}=1; $cost{2}{40}=1; $cost{2}{41}=1; $cost{2}{42}=1; $cost{2}{43}=1; $cost{2}{44}=1; $cost{2}{45}=1; $cost{2}{46}=1; $cost{2}{47}=1; $cost{2}{48}=1; $cost{2}{49}=1; $cost{2}{50}=1;

Flows within "Group" (2 means flow between 3 replacement nodes):
$cost{3}{4}=1; $cost{3}{5}=1;
$cost{4}{3}=1; $cost{4}{10}=1;
$cost{5}{3}=1; $cost{5}{6}=2; $cost{5}{7}=2;
$cost{6}{5}=2; $cost{6}{7}=2; $cost{6}{8}=1;
$cost{7}{5}=2; $cost{7}{6}=2; $cost{7}{13}=1;
$cost{8}{6}=1; $cost{8}{9}=1;
$cost{9}{8}=1; $cost{9}{15}=1;
$cost{10}{4}=1; $cost{10}{11}=2; $cost{10}{12}=2;
$cost{11}{10}=2; $cost{11}{12}=2; $cost{11}{18}=1;
$cost{12}{10}=2; $cost{12}{11}=2; $cost{12}{25}=1;
$cost{13}{7}=1; $cost{13}{14}=1;
$cost{14}{13}=1; $cost{14}{20}=1;
$cost{15}{9}=1; $cost{15}{16}=2; $cost{15}{17}=2;
$cost{16}{15}=2; $cost{16}{17}=2; $cost{16}{23}=1;
$cost{17}{15}=2; $cost{17}{16}=2; $cost{17}{27}=1;
$cost{18}{11}=1; $cost{18}{19}=1;
$cost{19}{18}=1; $cost{19}{21}=1;
$cost{20}{14}=1; $cost{20}{21}=2; $cost{20}{22}=2;
$cost{21}{19}=1; $cost{21}{20}=2; $cost{21}{22}=2;
$cost{22}{20}=2; $cost{22}{21}=2; $cost{22}{24}=1;
$cost{23}{16}=1; $cost{23}{24}=1;
$cost{24}{22}=1; $cost{24}{23}=1;
$cost{25}{12}=1; $cost{25}{26}=1;
$cost{26}{25}=1; $cost{26}{38}=1;
$cost{27}{17}=1; $cost{27}{28}=1;
$cost{28}{27}=1; $cost{28}{43}=1;
$cost{29}{30}=1; $cost{29}{31}=1;
$cost{30}{29}=1; $cost{30}{36}=1;
$cost{31}{29}=1; $cost{31}{32}=2; $cost{31}{33}=2;
$cost{32}{31}=2; $cost{32}{33}=2; $cost{32}{34}=1;
$cost{33}{31}=2; $cost{33}{32}=2; $cost{33}{39}=1;
$cost{34}{32}=1; $cost{34}{35}=1;
$cost{35}{34}=1; $cost{35}{41}=1;
$cost{36}{30}=1; $cost{36}{37}=2; $cost{36}{38}=2;
$cost{37}{36}=2; $cost{37}{38}=2; $cost{37}{44}=1;
$cost{38}{26}=1; $cost{38}{36}=2; $cost{38}{37}=2;
$cost{39}{33}=1; $cost{39}{40}=1;
$cost{40}{39}=1; $cost{40}{46}=1;
$cost{41}{35}=1; $cost{41}{42}=2; $cost{41}{43}=2;
$cost{42}{41}=2; $cost{42}{43}=2; $cost{42}{49}=1;
$cost{43}{28}=1; $cost{43}{41}=2; $cost{43}{42}=2;
$cost{44}{37}=1; $cost{44}{45}=1;
$cost{45}{44}=1; $cost{45}{47}=1;
$cost{46}{40}=1; $cost{46}{47}=2; $cost{46}{48}=2;
$cost{47}{45}=1; $cost{47}{46}=2; $cost{47}{48}=2;
$cost{48}{46}=2; $cost{48}{47}=2; $cost{48}{50}=1;
$cost{49}{42}=1; $cost{49}{50}=1;
$cost{50}{48}=1; $cost{50}{49}=1;

Costs to "O":

$cost{3}{51}=3;  $cost{4}{51}=3;  $cost{5}{51}=3;  $cost{6}{51}=3;  $cost{7}{51}=3;
$cost{8}{51}=3;  $cost{9}{51}=3;  $cost{10}{51}=3;  $cost{11}{51}=3;
$cost{12}{51}=3;  $cost{13}{51}=3;  $cost{14}{51}=3;  $cost{15}{51}=3;
$cost{16}{51}=3;  $cost{17}{51}=3;  $cost{18}{51}=3;  $cost{19}{51}=3;
$cost{20}{51}=3;  $cost{21}{51}=3;  $cost{22}{51}=3;  $cost{23}{51}=3;
$cost{24}{51}=3;  $cost{25}{51}=3;  $cost{26}{51}=3;  $cost{27}{51}=3;
$cost{28}{51}=3;  $cost{29}{51}=3;  $cost{30}{51}=3;  $cost{31}{51}=3;
$cost{32}{51}=3;  $cost{33}{51}=3;  $cost{34}{51}=3;  $cost{35}{51}=3;
$cost{36}{51}=3;  $cost{37}{51}=3;  $cost{38}{51}=3;  $cost{39}{51}=3;
$cost{40}{51}=3;  $cost{41}{51}=3;  $cost{42}{51}=3;  $cost{43}{51}=3;
$cost{44}{51}=3;  $cost{45}{51}=3;  $cost{46}{51}=3;  $cost{47}{51}=3;
$cost{48}{51}=3;  $cost{49}{51}=3;  $cost{50}{51}=3;

Cost to close loop (if one defines TSP as requirement for whole loop):
$cost{51}{1}=1;

All other costs should be considered as significantly large – for example 200.

## B. $x_{i,s,j}$ flow in counter example

[Picture, due to its size, is available in version:
http://www.teycom.pl/docs/Report_on_article_The_Travelling_Salesman_Problem_A_Linear_Programming_Formulation.pdf]

Figure 7 Counter example flow

## C. Algorithm for obtaining CE flow

```
solution_x{"x_1_1_2"}=$total_flow_constant;
for (my $j=3;$j<=50;$j++)
{
  $solution_x{"x_2_2_$j"}=$total_flow_constant/48;
}
for (my $s=3;$s<50;$s++)
{
  for (my $i=3;$i<=50;$i++)
  {
    # for cost == 1 - flow = 1/2, for cost == 2 flow == 1/4
    for (my $j=3;$j<=50;$j++)
    {
      next if ($i==$j);
      if (get_cost($i,$j)==1)
      {
        $solution_x{"x_$i\_$s\_$j"}=($total_flow_constant/48)/2;
      }
      elsif (get_cost($i,$j)==2)
      {
        $solution_x{"x_$i\_$s\_$j"}=($total_flow_constant/48)/4;
      }
    }
  }
}
for (my $i=3;$i<=50;$i++)
{
  $solution_x{"x_$i\_50_51"}=$total_flow_constant/48;
}
$solution_x{"x_51_51_1"}=$total_flow_constant;
```